\documentclass[aps,prd,12pt,notitlepage,nofootinbib,tightenlines]{revtex4-1}
\usepackage{amsmath}
\usepackage{bm}
\usepackage{times}
\usepackage{braket}
\usepackage{color}
\usepackage{epsfig}
\usepackage{slashed}
\usepackage{hyperref}
\newcommand{\beq}{\begin{eqnarray}}
\newcommand{\eeq}{\end{eqnarray}}
\newcommand{\non}{\nonumber\\ }
\newcommand{\ov}{\overline}

\newcommand{\psl}{ P \hspace{-2.8truemm}/ }

\def\lsim{ {\ \lower-1.2pt\vbox{\hbox{\rlap{$<$}\lower6pt\vbox{\hbox{$\sim$}}}}\ } }
\def\gsim{ {\ \lower-1.2pt\vbox{\hbox{\rlap{$>$}\lower6pt\vbox{\hbox{$\sim$}}}}\ } }
\def \epjc{Eur Phys J C}

\def \jpg{J Phys G}
\def \npb{Nucl Phys B}

\def \plb{Phys Lett B}

\def \prd{Phys Rev D}
\def \prl{Phys Rev Lett}
\def \zpc{Z Phys C}
\def \jhep{J High Energy Phys}

\def \rmp{Rev Mod Phys}
\def \ppnp{Prog Part $\&$ Nucl Phys}

\definecolor{Red}{rgb}{1.,0.,0.}

\definecolor{Blue}{rgb}{0.,0.,1.}

\definecolor{nicered}{rgb}{0.7,0.1,0.1}
\definecolor{nicegreen}{rgb}{0.1,0.5,0.1}

\hypersetup{colorlinks,citecolor=nicegreen,linkcolor=nicered}

\begin{document}

\title{ The two-body hadronic decays of $B_c$ meson in the perturbative QCD approach: A short review}
\author{Zhen-Jun Xiao$^{1,2}$}
\email[Electronic address:]{xiaozhenjun@njnu.edu.cn}
\affiliation{1. Department of Physics and Institute of Theoretical
Physics, Nanjing Normal University, Nanjing, Jiangsu 210023,
People's Republic of China}
\affiliation{2. Jiangsu Key Laboratory for Numerical Simulation of Large Scale Complex Systems,
Nanjing Normal University, Nanjing 210023, People's Republic of China,}
\author{Xin Liu}
\email[Electronic address:]{liuxin.physics@gmail.com}
\affiliation{School of Physics and Electronic Engineering, Jiangsu Normal University, Xuzhou, Jiangsu 221116, People's Republic of China}

\date{\today}

\begin{abstract}

\noindent{\bf \large \hspace{-0.8cm} Abstract} \ \ \ \
Along with the running of Large Hadron Collider (LHC) located at CERN in November 2009,
a large number of data samples of $B_c$ meson have been collected and some
hadronic $B_c$ decay modes have been measured by the LHC experiments.
In view of the special and important roles of $B_c$ meson decays playing in the
heavy flavor sector, we here give a short review on the status of two body
hadronic decays $B_c \to M_1 M_2$  at both experimental and theoretical aspects.
For the theoretical progresses, specifically, we will show lots of theoretical studies on two body
hadronic $B_c$ decays involving pseudoscalar, vector, scalar, axial-vector, even tensor meson(s)
in the final states by employing the perturbative QCD (pQCD) factorization approach.
We will present a general analysis about the two-body hadronic decays of the heavy $B_c$ meson
and also provide some expectations for the future developments.

\end{abstract}

\pacs{13.25.Hw, 12.38.Bx, 14.40.Nd}

\maketitle
{\bf Key Words}\hspace{0.5cm}  $B_c$ meson hadronic decays; The pQCD factorization approach;
           Branching ratios;  CP-violating asymmetries; Polarization fraction

%
%

\section*{1. Introduction}\label{sec:intro}

The $B_c$ meson is the lowest-lying bound state of $\bar{b}$ and $c$ quark
with $J^P = 0^-$ in the standard model(SM)~\cite{Abulencia2006:bc}.
It is too heavy to be produced in the old $B$ factories at KEK and SLAC, but it can be produced in
significant numbers in high energy hadron collisions, such as the Tevatron and LHC experiments.
The heavy $B_c$ meson was first discovered by CDF collaboration at Tevatron in 1998 through the
semileptonic modes $B_c \to J/\psi(\mu^+\mu^-) l^+ X(l = e, \mu)$~\cite{Abe1998}, which
demonstrated the possibility for investigations on $B_c$ physics experimentally.
At the Large Hadron Collider (LHC) experiments, furthermore, a large number of $B_c$ meson
events could be collected. With a luminosity of about ${\cal L}= 10^{34} {\rm cm}^{-2} {\rm s}^{-1}$,
around $5 \times 10^{10}$ $B_c$ events are expected to be produced each year~\cite{Kiselev2004:bc}.
The properties of $B_c$ meson and the dynamics involved in the $B_c$ decays would be fully exploited
through the precision measurements at the LHC with its high collision energy and high luminosity.
A golden era of $B_c$ physics is opened with the successful running of the LHC experiments, especially
the measurements carried on by the LHCb Collaboration, where about $1\%$ of the total $b$-related data sample
are the $B_c$ events: $10^{9}\sim 10^{10}$ $B_c$ decays each year.

\begin{figure}[!!htb]
  \centering
  \begin{tabular}{l}
  \includegraphics[width=0.45\textwidth]{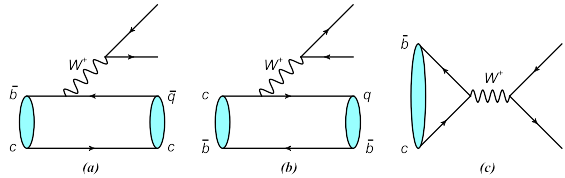}
  \end{tabular}
  \caption{Typical Feynman diagrams for three types   of $B_c$ decays:
  (a) $\bar{b}$ weak decay  modes with $q= c$ or $u$,
  (b) $c$ weak decay modes with $q= s$ or $d$,
  and (c) pure weak annihilation channels, respectively.}
  \label{fig:fig1}
\end{figure}

The $B_c$ meson is unique because it is flavor-asymmetric, which is very different
from the symmetric heavy quarkonium states, i.e., $c \bar c$ and $b \bar b$. It is the only weakly decaying doubly
heavy flavor meson since the two flavor-asymmetric quarks($b$ and $c$) cannot annihilate into gluons or photons via
strong interactions or electromagnetic interactions, which offers a novel window for studying the heavy quark dynamics
that is inaccessible through the investigations on the $b\bar b$ and $c\bar c$ quarkonia.
The $B_c$ meson is expected to decay through the weak interaction and has rich decay channels that could
provide an ideal platform to study hadronic weak decays of heavy quark
flavor~\cite{Brambilla2011:bc}
in the SM. The decay processes of the heavy $B_c$ meson can be subdivided into three types as follows~\cite{Ji2010:bc-phd}:
\begin{itemize}

\item[(1)]
$\bar b$ weak decay modes: $\bar b \to (\bar{c}, \bar{u}) W^+$, which will result in
the final states such as $J/\psi l \bar{\nu}_l$, $J/\psi \pi^+$, etc., as shown
in Fig.~\ref{fig:fig1}(a);

\item[(2)]
$c$ weak decay modes: $c \to (s, d) W^+$, which will lead to the final states such
as $B_s l \bar{\nu}_l$, $B_s \pi^+$, etc., as shown in Fig.~\ref{fig:fig1}(b);

\item[(3)]
pure weak annihilation channels: $\bar b c \to W^+$, which will give the
final states such as $B_c \to l \bar{\nu}_l$, $\overline{K}^{(*)0} K^{(*)+}$,
etc., as illustrated in Fig.~\ref{fig:fig1}(c) .

\end{itemize}

From a theoretical point of view,
the weak hadronic decays of $B_c$ meson are extremely complicated due to
its heavy-heavy nature and the participation of strong interaction, which
complicate the extraction of  parameters
in the SM, but they also provide great opportunities to study the perturbative
and nonperturbative QCD, final state interactions, and heavy quarkonium properties, etc.
So far, to our best knowledge, lots of hadronic $B_c$ decays have been studied
extensively within various of theoretical approaches/methods
in the literature, for example in
Refs.~\cite{Hussain1984:bc,Du1989:bc,Masetti1992:bc,Xu1992:bc,Chang1994:bc,
Gershtein1995:bc,Sanchis-Lozano1995:bc,Kiselev1996:bc,
Du1996:bc,Liu1997:bc,Du1998:bc,Du1999:bc,El-Hady2000:bc,
Fleischer2000:bc,Guo2001:bc,Saleev2000:bc,Saleev2001:bc,Giri2002:bc,Verma2002:bc,
Kiselev2002:bc,Castro2002:bc,Ivanov2003:bc,Ebert2003:bc-1,
Ebert2003:bc-2,Kiselev2004:bccp,Fajfer2004:bc,Ivanov2006:bc,
Hernandez2006:bc,Giri2007:bc,Wang2007:ncbc,Dhir2008:bc,
Sun2008:bc-1,Sun2008:bc-2,Liu2008:bc,Sun2009,Cheng2006:bcdpi,
Zhang2009:bcdk,Choi2009:bc,D-Genon2009:ncbc,Rakitin2010:ncbc,
Likhoded2010:ncbc,Sharma2010:bc-1,Sharma2010:bc-2,Sharma2010:bc-3,
Liu2010:bcpv,Liu2010:bcap,Yu2010:bcpsik,Yang2010:bckk,Liu2010:bcspv,
Ebert2010:bc,Fu2011:bc,Wang2012:bc,Liu2011:bcav,Xiao2011:bcaa,Zhou12:bcdpv,
Zou13:bcdt,Zhou12:bcdd,wang14a,Wang2012:ncbc,Luchinsky2012:ncbc,
Qiao2012:ncbc,Naimuddin2012:ncbc,Dhir2013:ncbc-1,Dhir2013:ncbc-2,
Esposito2013:ncbc,Kar2013:ncbc}.

At the quark level, the effective weak Hamiltonian
$H_{\rm eff} \propto \sum C_i(\mu) O_i(\mu)$ is theoretically
well under control, where $O_i$ are local four-quark operators and $C_i(\mu)$ are the
Wilson coefficients which incorporate strong-interaction effects above the scale
$\mu$. However, it is a difficult task to evaluate the hadronic matrix elements
of $O_i$ reliably due to the nonperturbative QCD effects involved.
Since the $B_c$ meson is heavy, it is possible to describe the dynamics of hadronic decays
by theories motivated by QCD.
A  central aspect of those theories is the factorization theorem which allows one to
disentangle the short-distance QCD dynamics from the non-perturbative hadronic effects.
During the past decades, theorists have made great efforts on the evaluations of
hadronic matrix elements based on the QCD dynamics.
So far, the QCD factorization(QCDF) approach~\cite{Beneke99,Du02},
the soft-collinear effective theory(SCET)~\cite{Bauer04:scet}
and the perturbative QCD(pQCD) approach~\cite{Li01:kpi,Lu01:pipi,Li03:ppnp},
have been developed to make effective evaluations of hadronic matrix elements.
Furthermore, up to now, the well-defined pQCD approach~\cite{Li03:ppnp}
has become one of the most popular methods in the market due to its unique features
~\cite{Li03:ppnp}.

In this short review, we give an overview of the experimental measurements
and the theoretical understanding of the branching ratios and CP-violating
asymmetries of the two body hadronic $B_c\to M_1 M_2$ decays (here $M_i$ denotes various
mesons).
We begin with a brief summary on current status about the experimental measurements of the
hadronic $B_c$ decays. This is followed by an introduction to the
theories for the study of hadronic $B_c$ decays, and a discussion on the choice
of wave functions for doubly heavy flavor $B_c$ meson and hadrons involved
in the final states. Last but not least, we present some recent investigations
for the two body hadronic $B_c$ decays by employing the pQCD 
approach  at leading order and leading power.
Few of the pQCD predictions for the considered $B_c$ decays
have been tested now in the experiments, but some of them will be  measured
soon in the LHCb experiments. Finally, we make conclusions and a short summary.

%
\section*{2.  Hadronic $B_c$ decays: Experiments}\label{sec:exp-bc}

Before the running of the LHC at CERN, ever since the $B_c$ meson
was discovered by the CDF experiment at the Tevatron~\cite{Abe1998}, only one hadronic
decay mode of $B_c$ meson had been observed, $B_c \to J/\psi \pi^+$, which was utilized by
CDF and D0 Collaboration ~\cite{Aaltonen2008,cdf08} to measure the $B_c$ mass. The mass and
lifetime  of $B_c$ meson as given in Particle Data Group 2012\cite{pdg2012}
are the following:
\beq
m_{B_c} &=& (6274.5 \pm 1.8) {\rm MeV}, \non
\tau_{B_c} &=& (0.452\pm 0.033) {\rm ps}.\; \label{eq:mbc}
\eeq

Although the CMS and ATLAS Collaboration  reported their  observation of some $B_c$ decays,
such as $B_c^+ \to J/\Psi \pi^+$ \cite{atlas}, most $B_c$-related measurements have
been done by LHCb collaboration.
In Table \ref{tab:exp2013}, we list currently available data for
the relative branching ratios of hadronic decays of $B_c$ meson
and some other physical observables as reported by LHCb Collaboration
in Refs.~\cite{lhcb2013,lhcbtalk}.

\begin{table}[tb]
\begin{center}
\caption{ The measurements for some hadronic $B_c$ meson decays
as reported by LHCb Collaboration \cite{lhcb2013,lhcbtalk}. }
\label{tab:exp2013}
\vspace{0.2cm}
\begin{tabular}{l | c  } \hline \hline
Measured values of physical observables & Data  \\ \hline
$\frac{Br(B_c \to J/\psi \pi^+ \pi^- \pi^+)}{Br(B_c \to J/\psi \pi^+)}= 2.41 \pm 0.30 \pm 0.33$
& $0.8$ {\rm fb}$^{-1}$  \\ \hline
$R_{c/u} \equiv \frac{\sigma(B_c)Br(B_c \to J/\psi \pi^+)}{\sigma(B_u)Br(B_u
\to J/\psi K^+)} =  0.68 \pm 0.12$& $0.37$ {\rm fb}$^{-1}$ \\ \hline
$\frac{Br(B_c \to \psi(2S) \pi^+)}{Br(B_c \to J/\psi \pi^+)} =
0.250 \pm 0.068 \pm 0.014$ & $1.0$ {\rm fb}$^{-1}$  \\ \hline
$R_{D_s/\pi} \equiv \frac{Br(B_c \to J/\psi D_s^+)}{Br(B_c \to J/\psi \pi^+)}
 = 2.90 \pm 0.57 \pm 0.24$ & $3$ {\rm fb}$^{-1}$ \\ \hline
$R_{D_s^*/D_s} \equiv \frac{Br(B_c \to J/\psi D_s^{*+})}{Br(B_c \to J/\psi D_s^+)}
 = 2.37 \pm 0.56 \pm 0.10$& $3$ {\rm fb}$^{-1}$  \\ \hline
$f_{\pm\pm} \equiv \frac{Br_{\pm\pm}(B_c \to J/\psi D_s^{*+})}{Br(B_c \to J/\psi D_s^{*+})}
 = (52 \pm 20) \%$ &$3$ {\rm fb}$^{-1}$ \\ \hline
$\frac{Br(B_c \to J/\psi K^+)}{Br(B_c \to J/\psi \pi^+)} =
0.069 \pm 0.019 \pm 0.005$&$1.0$ {\rm fb}$^{-1}$  \\ \hline
$\frac{\sigma(B_c)}{\sigma(B_s)} \times Br(B_c \to B_s \pi^+) =
2.37^{+0.37}_{-0.35}\cdot 10^{-3}$&$3$ {\rm fb}$^{-1}$  \\ \hline
$\frac{Br(B_c \to J/\psi K^+ K^- \pi^+)}{Br(B_c \to J/\psi \pi^+)} =0.53 \pm 0.10\pm 0.05$&$3$ {\rm fb}$^{-1}$ \\ \hline
\hline
\end{tabular}
\end{center} \end{table}


In the following years, more and more hadronic decay modes of $B_c$ meson
will be measured with good  precision in the LHCb experiments.
Meanwhile, the theoretical predictions for the hadronic $B_c$
meson decays in various approaches/methods will be greatly
required in order to understand the measured results from the LHC experiments.

\section*{ 3.  Factorization approaches in the framework of QCD}\label{sec:3}

In this section, we will introduce the QCD-based factorization approaches/methods that have been
adopted for studying the dynamics of hadronic $B_c \to M_1 M_2$ decays.
It is worth of stressing that although charm is a ``heavy" quark, the mass around 1.5 GeV
makes the studies of $c \to (d,s)$ decays suffer from rather large long-distance
contributions and/or final state interactions, and
consequently makes the estimates of the relevant physical observables in
$B_c \to B_q X$ decays with  $q=(d,s)$ less trustworthy.
In fact, there are no any reliable predictions for the hadronic $B_c \to B_q X $ decays
based on the QCD-motivated factorization framework at present.
Therefore, we will not consider the $B_c \to B_q X$ decay modes in this paper.
We here will study the two body hadronic $B_c$ decays arising from the $\bar{b}$ decays
or the pure annihilation processes, as shown in Figs.~\ref{fig:fig1}(a) and
\ref{fig:fig1}(c).

In the effective Hamiltonian approximation, the decay amplitude of the considered hadronic
$B_c \to M_1 M_2$ decays can be written as
\beq
{\cal A}(B_c \to M_1 M_2) = < M_1 M_2| H_{\rm eff} | B_c >,\ \
\eeq
where $H_{\rm eff}$ is the corresponding weak effective Hamiltonian ~\cite{Buras96}
\beq
H_{\rm eff}\, &=&\, \frac{G_F}{\sqrt{2}}
\biggl\{ \sum_{Q=u,c} V^*_{Qb}V_{Qq} [ C_1(\mu)O_1^{Q}(\mu)
+C_2(\mu)O_2^{Q}(\mu) ]\non
&&
 - V^*_{tb}V_{tq} [ \sum_{i=3}^{10}C_i(\mu)O_i(\mu) ] \biggr\}+ {\rm H.c.}\;,
\label{eq:heff}
\eeq
with $q= (d,s)$, the Fermi constant $G_F=1.16639\times 10^{-5}{\rm
GeV}^{-2}$, CKM matrix elements $V_{ij}$, and Wilson coefficients $C_i(\mu)$ incorporating
strong-interaction effects above the scale $\mu$. The local four-quark
operators $O_i(i=1,\cdots,10)$ include the current-current(tree) operators
$O_{1,2}$, the QCD penguin operators $O_{3-6}$ and the electroweak penguin operators
$O_{7-10}$ \cite{Buras96}.

The key point in the theoretical calculations for the decay amplitude is how to evaluate the hadronic matrix elements
of the four-quark operators $<O_i>=<M_1 M_2| O_i| B_c>$ reliably.
Presently, there are three popular factorization approaches:
the QCDF approach\cite{Beneke99}, the SCET\cite{Bauer04:scet}
and the pQCD approach\cite{Li01:kpi,Lu01:pipi}.
A detailed discussion for these theories goes beyond the scope of this short review, and the interested
reader is referred to the original literatures.
Basically, theories of hadronic $B_c$ decays are based on the ``factorization theorem" under which the short-distance contributions
to the decay amplitudes can be separated from the process-independent long-distance parts.

\begin{figure*}
\centering
\includegraphics[width=0.7\textwidth]{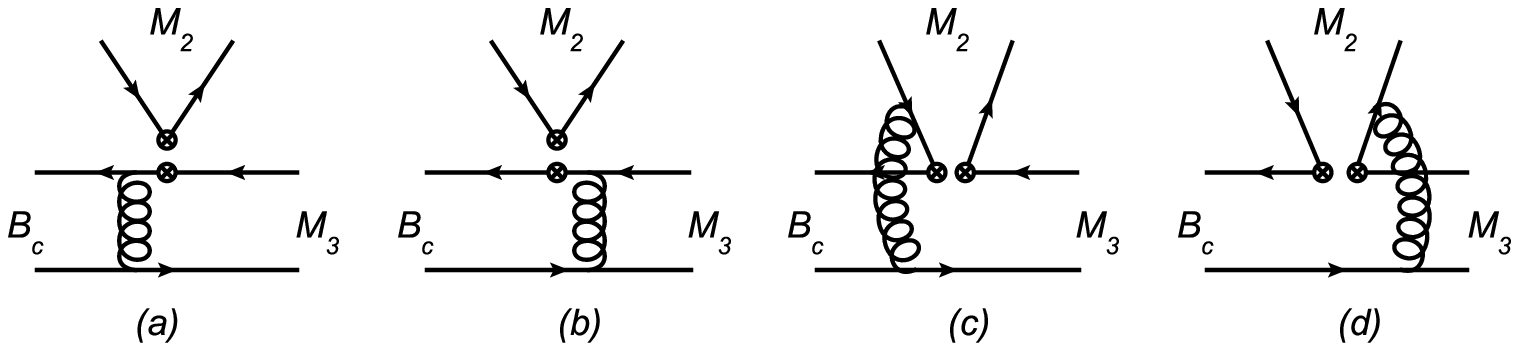}\\
\includegraphics[width=0.7\textwidth]{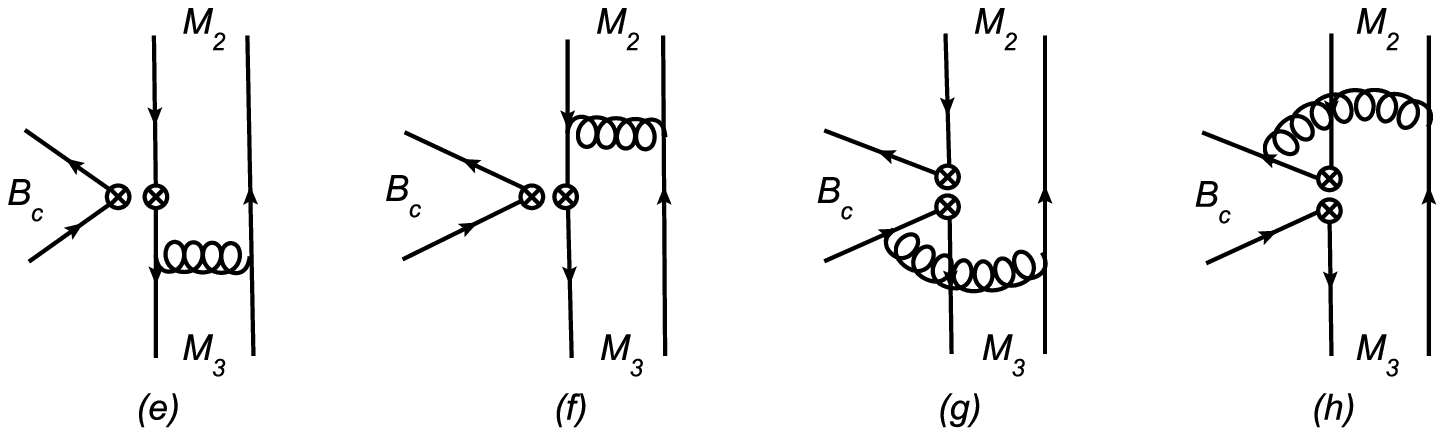}
\caption{Typical Feynman diagrams for $B_c\to M_1M_2$ decays at leading order
in the pQCD approach.}
\label{fig2}
\end{figure*}

In the process of calculating the hadronic matrix elements of $B_c$ meson decays,
we need to cope well with the physical scales around the so-called factorization
scale, namely, $\sqrt{\Lambda_{\rm QCD} m_b}$.
Usually, scales below this factorization scale are treated as the nonperturbative
physics, which is described by the transition form factors or hadron wave functions. Scales above
this factorization scale are categorized as the perturbative physics, which can be evaluated as the
expansion of the strong coupling constant $\alpha_s$ with various approaches/methods.

Both of the QCDF approach and the SCET are within the framework of collinear factorization.
In the QCDF approach,  the endpoint singularity appears at high twist calculations and the annihilation
type diagrams. Those annihilation types of diagrams are later proved to be important.
The SCET also leave part of the soft contribution in the form factor
diagrams as nonperturbative inputs, which make it less predictive, since it requires more free parameters
to be determined by experiments~\cite{Wang2008:scet}.
The predictions of the annihilation contributions in SCET are almost real with tiny strong
phase~\cite{Arnesen08:anni-scet}, which is rather different from almost imaginary with large strong
phase~\cite{Chay08:complexanni} as predicted in the pQCD approach.


The basic idea of the pQCD approach is that it takes into
account the transverse momentum $k_T$ of the valence quarks to kill the endpoint
divergence
in the calculation of the hadronic matrix elements. Therefore,
we have one more scale, i.e.,
the quark transverse momentum than the QCDF approach and the SCET.
The b-flavor meson transition form factors, and the spectator and
annihilation contributions are then all calculable in the framework
of the $k_T$ factorization, where three energy scales are
involved~\cite{Li01:kpi,Lu01:pipi,Li95:pQCD}.
The hard dynamics is characterized by $\sqrt{m_b \Lambda_{\rm QCD}}$,
which is to be perturbatively calculated.
The harder dynamics is from $m_W$ scale to $m_B$ scale described by
renormalization group equation
for the four quark operators. The dynamics below
$\sqrt{m_b \Lambda_{\rm QCD}}$ is soft, which is described by the meson wave
functions. The soft dynamics is not perturbative but universal for all channels.
In the pQCD approach, a $B_c \to M_1 M_2$
decay amplitude is therefore factorized into the convolution of the six-quark
hard kernel($H$), the jet function($J$) and the Sudakov factor($S$) with the
bound-state wave functions($\Phi$) as follows,
\begin{eqnarray}
{\cal A}(B_c \to M_1 M_2)=\Phi_{B_c} \otimes H \otimes J \otimes S
\otimes \Phi_{M_1} \otimes \Phi_{M_2},\ \  \label{eq:sixquarks}
\end{eqnarray}
All nonperturbative components are organized in the form of hadron wave
functions $\Phi$, which may be extracted from experimental data or
other nonperturbative method, such as QCD sum
rules~\cite{QCDSR}. Since nonperturbative dynamics has been factored out,
one can evaluate all possible Feynman diagrams presented in
Fig.~2 for the six-quark amplitude $H$ straightforwardly.
The jet function $J$ comes from the threshold resummation, which
exhibits suppression in the small $x$ (quark momentum fraction)
region~\cite{Li02:threshold}. The
Sudakov factor $S$ comes from the $k_T$ resummation \cite{Sterman92,huang91},
which exhibits suppression in the small $k_T$ region.
Therefore, these resummation
effects guarantee the removal of the endpoint singularities and the reliability of the pQCD approach.

\section*{4.  Relevant hadron wave functions}\label{ssec:hadr-wf}

In order to calculate the analytic formulas of the decay amplitudes, we need the light cone wave functions decomposed
in terms of the spin structure. In general, the light cone wave functions are decomposed into 16 independent
components, $1_{\alpha\beta}$, $\gamma^{\mu}_{\alpha\beta}$, $\sigma^{\mu\nu}_{\alpha\beta}$,
$(\gamma^{\mu}\gamma_5)_{\alpha\beta}$, and $\gamma_{5\alpha\beta}$.
Relative to the more heavier $b$ quark, charm can be viewed approximately as a light quark in the doubly heavy flavor $B_c$ meson.
In the leading order of $m_c/m_{B_c} \sim 0.2$ expansion, the factorization theorem is
applicable to the $B_c$ system similar to the situation of $B$ meson.
In analogy to the definition of the $B$ meson~\cite{Li01:kpi,Lu01:pipi}, the light-cone wave function of $B_c$ meson
can be defined as
\beq
\Phi_{B_c,\alpha\beta,ij}&\equiv&
  \langle 0|\bar{b}_{\beta j}(0)c_{\alpha i}(z)|B_c(P)\rangle \non
&=& \frac{i \delta_{ij}}{\sqrt{2N_c}}\int dx d^2 k_T e^{-i (xP^-z^+ - k_T z_T)}\non
&&\cdot \Bigl \{ (\psl +m_{B_c})\gamma_5  \phi_{B_c}(x, k_T) \Bigr \}_{\alpha\beta}\;;
\label{eq:def-bq}
\eeq
where the indices $i,j$ and $\alpha,\beta$ are the Lorentz indices and color indices, respectively,
$P(m)$ is the momentum(mass) of the $B_c$ meson, $N_c$ is the color factor, and
$k_T$ is the intrinsic transverse momentum
of the lighter quark in $B_c$ meson. Note that, in principle, there are two Lorentz
structures of the wave function to be considered in the numerical calculations, however,
the contribution induced by the second Lorentz structure
is numerically small and approximately negligible.

In Eq.~(\ref{eq:def-bq}), $\phi_{B_c}(x,k_T)$ is the $B_c$ meson distribution amplitude
and obeys to the normalization condition:
$\int_0^1 dx\; \phi_{B_c}(x, b=0) = f_{B_c}/(2 \sqrt{2N_c})$, here $b$ is the
conjugate space coordinate of transverse momentum $k_T$ and $f_{B_c}$
is the decay constant of $B_c$ meson.

To our best knowledge, however, $\phi_{B_c}$ with $k_T$ is still absent now.
But, the situation may become somewhat simpler, if $B_c$ meson
can be approximated as a non-relativistic bound state of two sufficiently heavy quarks.
In this sense we expect exclusive matrix elements, in
particular, the light-cone distribution amplitude to be calculable perturbatively, since the quark masses provide
an intrinsic physical infrared regulator. At the nonrelativistic scale, the
leading 2-particle distribution amplitude can be approximated by
delta function, fixing the light-cone momenta of the quarks according to their masses~\cite{Bell2008:das-bc}.
Since $B_c$ meson consists of two heavy quarks and $m_{B_c} \simeq m_b+m_c$, the
distribution amplitude $\phi_{B_c}$ would thus be close to
$\delta(x-m_c/m_{B_c})$ in the non-relativistic limit. We therefore
adopt the non-relativistic approximation form of $\phi_{B_c}$ as~\cite{Bell2008:das-bc}
\beq
\phi_{B_c}(x) &=& \frac{f_{B_c}}{2 \sqrt{2 N_c}} \delta (x-
m_c/m_{B_c})\;,
\eeq
where the value of $f_{B_c}$ is taken from the calculations in quenched lattice QCD with
exact chiral symmetry~\cite{Chiu2007:fbc}.

For the final state wave functions, such as pseudoscalar, vector, scalar, axial-vector, even tensor mesons, we
refer the readers to the papers dealing with various
decay channels, for example, in Refs.~\cite{Zou13:bcdt,Liu2010:bcpv,Liu2010:bcap}.
But, we should stress here that the $k_T$ dependence of the distribution amplitudes in the final states
has been neglected, since its contribution is very small as indicated in the
reference~\cite{Li95:pQCD}. The underlying reason is that the contribution
from $k_T$ correlated with a soft dynamics is strongly suppressed by the Sudakov
effect through resummation for the wave function, which is dominated
by a collinear dynamics.

\section*{5.  $B_c\to M_1 M_2$ decays } \label{sec:numr-bc}

In this section, we will summarize current status of the theoretical
studies for the hadronic  $B_c\to M_1 M_2$ decays by employing the pQCD factorization approach.
In order to make numerical evaluations one need the input parameters, such as
the relevant masses, decay constants and lifetimes etc.
Since some parameters change from time to time, one needs to consult the
original paper for specific input parameters for each predictions as given in different works.

\subsection*{ 5.1  $B_c$ decays through $b \to (c,u)$ transitions}\label{ssec:cspec-bc}

Up to now, the decays $B_c \to D (\pi, K)$~\cite{Cheng2006:bcdpi,Zhang2009:bcdk},
$B_c \to (J/\psi, \eta_c) (\pi, K)$~\cite{Yu2010:bcpsik,Sun2009},
$B_c \to D_{(s)}^{(*)} (P, V, T)$~\cite{Zhou12:bcdpv,Zou13:bcdt}
and $B_c \to (D^{(*)}, D_s^{(*)})(D^{(*)}, D_s^{(*)})$~\cite{Zhou12:bcdd}
have been studied in the pQCD approach, and the CP-averaged branching ratios
and CP-violating asymmetries for these decay modes have been calculated.
Since the charm quark in the heavy final state mesons
( for example,$J/\psi$, $\eta_c$, $D^{(*)}$ and $D_s^{(*)}$) is almost at collinear state,
a hard gluon is needed to transfer large momentum to the spectator charm quark.
Utilizing the $k_T$ factorization instead of collinear factorization, the pQCD approach
is free of endpoint singularity. Thus Feynman diagrams as illustrated in
Fig.2 are all contributing and calculable.

In $B_c$ decays, there is one more intermediate energy scale, the heavy charm mass.
As a result, another expansion series of $m_c/m_{B}$ will appear.
The factorization is approved at the leading order of $m_c/m_{B}$
expansion~\cite{Lu02:b2d,Keum04,Li08:b2d}.
The nonleptonic $B_c$ to $J/\psi(\eta_c)$ and a light hadron decays are
similar to that of $B$ decaying into $D$ and a light meson~\cite{Sun2009}.

The two body hadronic $B_c \to (J/\psi, \eta_c) (\pi, K)$ decays are tree dominated modes including factorizable emission
and nonfactorizable spectator amplitudes. Furthermore, the branching ratios for the considered decays are determined
by the contributions arising from the factorizable topologies.
The decay amplitudes for $B_c \to (J/\psi, \eta_c) (\pi, K)$ can thus be described approximately as
 \beq
 {\cal A} (B_c \to J/\psi(\eta_c)\; \pi) &=& V_{cb}^* V_{ud}\cdot f_\pi \non
 && \cdot \langle J/\psi(\eta_c) |V-A|B_c\rangle \;, \\
 {\cal A} (B_c \to J/\psi(\eta_c)\; K) &=& V_{cb}^* V_{us}\cdot f_K \non
 && \cdot \langle J/\psi(\eta_c) |V-A|B_c\rangle \;;
 \eeq
in which the former mode is CKM favored, while the latter channel is CKM suppressed,
and $f_\pi$ and $f_K$ are the decay constants of pion and kaon respectively.
Then the interesting relation of branching ratios of the considered four decay modes
 in the limit of SU(3) flavor symmetry can be read as
 \beq
 R_{J/\psi}^{K/\pi} &=& R_{\eta_c}^{K/\pi} \sim \left |\frac{V_{us}}{V_{ud}}\right |^2 \cdot
 \left |\frac{f_K}{f_\pi}\right |^2 \;;
 \eeq
where $R_{J/\psi}^{K/\pi}$ and $R_{\eta_c}^{K/\pi}$ are defined as
\beq
 R_{J/\psi}^{K/\pi} &\equiv& \frac{Br(B_c \to J/\psi K)}{Br(B_c \to J/\psi \pi)}\;, \non
 R_{\eta_c}^{K/\pi} &\equiv & \frac{Br(B_c \to \eta_c K)}{Br(B_c \to \eta_c \pi)}\;;
\eeq
With the input parameters~\cite{pdg2012}: $V_{ud} = 0.97427$, $V_{us} = 0.22534$, $f_K = 0.16$ GeV,
and $f_\pi = 0.13$ GeV, the expected ratios $R_{J/\psi}^{K/\pi} = R_{\eta_c}^{K/\pi} \approx 0.08$.
From Ref.~\cite{Yu2010:bcpsik}, we found that
\beq
Br(B_c \to J/\psi \pi)_{\rm pQCD}&=& (1.35 \sim 2.54) \times 10^{-3}, \non
Br(B_c \to J/\psi K)_{\rm pQCD} &=& (1 \sim 3)\times 10^{-4},
\eeq
which lead to the ratio $R_{J/\psi}^{K/\pi} = (0.07 \sim 0.12)$ in the pQCD approach,
which is consistent with the naive expectation on the above ratio.
Very recently, the LHCb Collaboration has measured the ratio of the branching
ratios between  $B_c \to J/\psi K$ and $B_c \to J/\psi \pi$ decays,
and obtained the result ~\cite{lhcbtalk},
\beq
R_{J/\psi}^{K/\pi} = 0.069 \pm 0.020,
\eeq
which is in good agreement with the theoretical prediction in the pQCD approach.

From the pQCD prediction of $Br(B_c \to \eta_c \pi)_{\rm pQCD} = (1.47 \sim 2.79) \times 10^{-3}$ ~\cite{Sun2009},
it is expected that the CP-averaged branching ratio of $B_c \to \eta_c K$ mode may be
$Br(B_c \to \eta_c K)_{\rm pQCD} = (1 \sim 3) \times 10^{-4}$, which will be tested by the forthcoming experiments.
Of course, since no penguin operators are involved in these considered four
channels, the direct CP asymmetries are absent here naturally.

The two body hadronic $B_c$ meson decaying into double charm hadrons~\cite{Zhou12:bcdd},
which are the pure tree decay modes, can be utilized particularly to extract the CKM angles because of the
absence of the interference from the penguin operators.
Furthermore, the decays $B_c \to D_s^+ D^0$ and $D_s^+ \bar{D}^0$ are the gold-plated
modes for the extraction of CKM angle $\gamma$ through amplitude relations because their decay widths are expected to be at
the same order in magnitude~\cite{Masetti1992:bc,Fleischer2000:bc,Giri2002:bc,Ivanov2003:bc,Kiselev2004:bccp,Giri2007:bc}.

From the numerical calculations one found that the ratio of the decay widths for
$B_c \to D_s^+ D^0$ and $B_c \to D_s^+ \bar{D}^0$ is about $1.3$,
which indicate that the branching ratios for these two decays are really as was expected and they are indeed
suitable for extracting the CKM angle $\gamma$.
The theoretical predictions as given in Ref.~\cite{Zhou12:bcdd} confirmed that
the nonfactorizable spectator diagrams provided a remarkable contribution
in the double charm decays of $B_c$ meson.
The predicted branching ratios for the considered decay channels vary in the range of $10^{-8} \sim 10^{-5}$.
The considered $B_c$ decays with a decay rate at the level of $10^{-6}$ or larger can be detected with a good precision
at LHC experiments~\cite{D-Genon2009:ncbc}. Meanwhile, the transverse polarization fractions of the $B_c$ meson
decays with two vector $D^*$ mesons are predicted for the first time in the pQCD approach.
The transverse polarization fractions are large
in some channels, which mainly come from the nonfactorizable spectator diagrams.

For the nonleptonic $B_c \to D_{(s)}^{(*)} (P, V)$ decays, their decay rates and CP-violating asymmetries,
as well as the transverse polarization fractions for $B_c \to D_{(s)}^* V$ channels
are calculated systematically in the pQCD approach in Ref.~\cite{Zhou12:bcdpv}.
From the numerical calculations, one finds that the pQCD predictions for the CP-averaged branching ratios
of the tree-dominant $B_c \to D_{(s)}^{(*)} (P, V)$ decays are in good
agreement with that in the relativistic constituent quark model~\cite{Liu1997:bc}.

Furthermore, it is found that the nonfactorizable spectator diagrams
and annihilation diagrams
have remarkable effects on the physical observables in many channels,
especially in the color-suppressed and annihilation-dominant decay modes.
As expected, the annihilation diagrams give large contributions in the
$B_c$ meson decays, because the
contributions arising from annihilation diagrams are enhanced by the CKM
factor $V_{cb}^* V_{cq}$. For the $b \to s$ transition process, the ratio
$|V_{cb}^* V_{cs}/ V_{ub}^* V_{us}| \approx 47$, which therefore results
in the ratio $Br(B_c \to D^{(*)} K^{(*)+})/
Br(B_c \to D^{(*)+} K^{(*)0}) \approx 1$ for the considered two kinds of
annihilation-dominant modes.

For $B_c \to D_{(s)}^* V$ decays, furthermore,
the transverse polarization contributions are usually suppressed by the factor
$r_V$ or $r_{D_{(s)}^*}$ when compared with the longitudinal part.
Thus, for tree-dominant $B_c \to D^{*0} \rho^+$
and pure penguin type $B_c \to D^{*+} \phi$ decays, one can find the
relatively small transverse polarization fractions
$16.4\%$ and $11.5\%$, respectively.

For other $B_c \to D_{(s)}^* V$ decays, the annihilation contributions
dominate the branching ratios due to the
large Wilson coefficients. Therefore, the transverse polarization contributions
take a larger ratio in the branching ratios, which can reach
$50\% \sim 70\%$.
Because of the different weak phase and strong phase from tree diagrams,
penguin diagrams, and annihilation
diagrams, the possibly large direct CP violation in some channels are
predicted in the pQCD approach, for example,
\beq
&& A_{\rm CP}^{\rm dir}(B_c \to D^+ \rho^0) \approx 79.8\%, \non
&& A_{\rm CP}^{\rm dir} (D^0 K^{*+}) \approx -66.2 \%,\; {\rm etc.}
\eeq

For the hadronic $B_c \to D^{(*)} T$ decays ~\cite{Zou13:bcdt},
which is slightly special compared with the above
$B_c \to D_{(s)}^{(*)} (P, V)$ decays, there are no contributions from
factorizable emission diagrams because the emitted
tensor meson cannot be generated from the (axial-)vector current or
(pseudo-)scalar density. Thus, these $B_c \to D^{(*)} T$ decays
are forbidden in the naive factorization. One should go beyond the
naive factorization to calculate the nonfactorizable
spectator and annihilation diagrams. What's more, the annihilation
amplitudes are dominant in these considered $B_c \to D^{(*)} T$
decays because they are proportional to the large CKM  matrix
elements $V_{cb}$ and $V_{cd(s)}$.

The predictions in Ref.~\cite{Zou13:bcdt} show that the CP-averaged branching ratios
for hadronic $B_c \to D^{(*)} T$ modes are in the range of  $10^{-4} \sim 10^{-9}$. As stated in
Ref.~\cite{lhcb2013,lhcbtalk}, the LHC experiments, specifically the LHCb experiment, can produce around
$5 \times 10^{10}$ $B_c$ events per year. The $B_c$ decays with a decay
rate at the level of $10^{-6}$ can be detected with a good precision at
LHC experiments. Therefore, it is of great interests that the $B_c$ meson decays to tensor final states
with branching ratios as large as $10^{-4}$,
for example, $B_c \to D^* K_2^*(1430)$ and $B_c \to D_s^{*+} f_2^{'}(1525)$, will
be easier for experiments to search than the corresponding decays with vector mesons.
The modes with large branching ratios such as $B_c \to D^0 {K_2^{*}(1430)}^{+}$, $D^+
{K_2^*}(1430)^0$, $D_s^+ f_2^\prime(1525)$, $D^* K_2^*(1430)$, and
$D_s^{*+} f_2^{'}(1525)$, would provide opportunities to study the
properties of $B_c$ meson and the factorization theorem in the decays with an emitted tensor meson.

Most of the direct CP asymmetries for $B_c \to D^{(*)} T$ decays predicted
in the pQCD approach are very small because the penguin contributions are
too small compared with the tree annihilation
contributions. The largest direct CP violation for $B_c \to D^{(*)} T$
decays estimated with the pQCD approach
is $18.2\%$, which belongs to the channel $B_c \to D^+ a_2(1320)^0$.

The predicted transverse polarization fractions for most annihilation-dominant
$B_c \to D^* T$ channels in the pQCD approach
are larger than $50\%$, except for two modes $B_c \to D^{*+} f_2^\prime(1525)$
with $R_T \sim 45.3\%$
and $B_c \to D_s^{*+} a_2(1320)^0$ with  $R_T \sim 12.7\%$ ~\cite{Zou13:bcdt}.
Moreover, it is very interesting to note
that the longitudinal polarization contributions in $B_c \to D_s^{*+} f_2(1270)$ only about $1.6\%$.

It is worth of mentioning that the semileptonic charmed decays $B_c^+ \to D^{(*)}_{(s)}(l^+\nu,l^+l^-,\nu\bar \nu )$
have been studied in the pQCD approach \cite{wang14a}. In Ref.~\cite{wang14a}, we studied
the semileptonic decays of $B_c^+\to D^{(*)}_{(s)}(l^+\nu,l^+l^-,\nu\bar\nu)$ (here $l$ stands for $e$, $\mu$ or $\tau$)
by using the relevant form factors $F_{0,+,T}(q^2)$, $V(q^2)$, $A_{0,1,2}(q^2)$ and
$T_{1,2,3}(q^2)$ for the $B_c^+ \to (D_{(s)},D^*_{(s)})$ transitions obtained by employing the pQCD
factorization approach. We calculated the decays rates for all considered semileptonic decays and found
numerically that (a) the relevant transition form factors obtained in this work
agree well with those from other methods;
(b) the size of the pQCD predictions for the branching ratios for the decays
with $b\to s$ or $b\to d$ transitions show clearly the effects of the CKM suppression;
and (c) the pQCD predictions for the ratios of the decay rates are
$R_D\approx 0.7$ and $R_{D^*}\approx 0.6$, which could be measured at LHCb soon.

\subsection*{ 5.2  Charmless hadronic $B_c\to PP,PV,VP,VV$ decays} \label{ssec:anni-bc1}

The charmless hadronic $B_c\to M_1 M_2$ decays (i.e. $M_i$ are the charmless light mesons)
can occur only via the weak annihilation diagrams in the 
SM. As discussed in Sec.~3, up to now,
the annihilation diagrams can be well treated only by employing the
pQCD approach due to its unique features.

Although there is a different
viewpoint on the evaluations of annihilation contributions proposed in the SCET,
the previous predictions on the annihilation contributions in heavy flavor
$B$ meson decays calculated with the pQCD approach
have already been tested at various aspects, for example, branching ratios of pure
annihilation $B_d \to D_s^- K^+$, $B_d \to K^+ K^-$, and $B_s \to \pi^+ \pi^-$
decays~\cite{Lu03:dsk,Li04:pippim,Lu07:bs2mm,Xiao11:pippim},
direct CP asymmetries of $B^0 \to \pi^+\pi^-$, $K^+\pi^-$
decays~\cite{Li01:kpi,Lu01:pipi,Hong06:direct}, and the
explanation of $B\to \phi K^*$ polarization
problem~\cite{Li05:kphi,Li05b}, which indicate that the
pQCD approach is a reliable method to deal with the annihilation diagrams.

By using the pQCD approach, the pure annihilation type of charmless
hadronic $B_c\to M_1 M_2$ decays, about 200 decay modes,
\beq
B_c &\to & PP, \quad PV, \quad VV, \quad AP, \quad AV, \non
&& AA, \quad SP, \quad SV,
\eeq
have been studied systematically in Refs.~\cite{Liu2010:bcpv,Yang2010:bckk,Liu2010:bcap,Liu2010:bcspv,
Liu2011:bcav,Xiao2011:bcaa}, where the term $S,P,V$ and $A$ refers to the scalar,
 pesuodo-scalar, vector and axial-vector charmless mesons respectively.
Other possible charmless $B_c\to M_1 M_2$ decays through pure annihilation topology
to light mesons with $M_iM_j=SS$, $SA$, $TP$, $TV$, $TS$, $TA$ and
even $TT$ are under study now by using the pQCD approach~\cite{Liu2014:bc-anni}.

For the twenty three charmless $B_c \to PP, PV/VP$ decays, the decay rate can be written as
\beq
\Gamma =\frac{G_{F}^{2}m^{3}_{B_c}}{32 \pi  } |{\cal A}(B_c
\to M_1 M_2)|^2\;
\eeq
Using the decay amplitudes as given in Eqs.(20)-(27) and (32)-(46) in Ref.~\cite{Liu2010:bcpv},
it is straightforward to calculate the branching ratios with uncertainties as listed in
Table \ref{tab:bcpp}.


\begin{table*}
\centering
\caption{The pQCD predictions of branching ratios for eight $B_c \to PP$ modes.
and fifteen $B_c \to (PV,VP)$ decays.
The dominant errors come from charm quark mass $\rm{m_c}=1.5 \pm 0.15$ GeV,
combined Gegenbauer moments $a_i$, and chiral enhancement factors $m_0^{\pi}=1.4 \pm 0.3$ GeV
and $m_0^{K}=1.6 \pm 0.1$ GeV, respectively.}
\label{tab:bcpp}
\begin{tabular}[t]{l|l|l|l} \hline  \hline
Decay Modes  &                & Decay Modes &  \\
$(\Delta S=0)$ & $BR's(10^{-8}$) & $(\Delta S=1)$ & $BR's(10^{-8}$) \\
\hline
$\rm{B_c \to \pi^+ \pi^0}$ &0&
$\rm{B_c \to \pi^+ K^0}$&$4.0^{+1.0}_{-0.6}(m_c)^{+2.3}_{-1.6}(a_i)^{+0.5}_{-0.3}(m_0)$
   \\
 $\rm{B_c \to \pi^+ \eta}$ &$22.8^{+6.9}_{-4.6}(m_c)^{+7.2}_{-4.5}(a_i)^{+3.4}_{-4.2}(m_0)$&
 $\rm{B_c \to K^+ \eta}$ & $0.6^{+0.0}_{-0.0}(m_c)^{+0.6}_{-0.5}(a_i)^{+0.2}_{-0.1}(m_0)$
 \\
 $\rm{B_c \to \pi^+ \eta'}$ & $15.3^{+4.6}_{-3.1}(m_c)^{+4.8}_{-3.0}(a_i)^{+2.2}_{-2.8}(m_0)$&
$\rm{B_c \to K^+ \eta'}$ & $5.7^{+0.9}_{-0.9}(m_c)^{+1.0}_{-1.6}(a_i)^{+0.0}_{-0.3}(m_0)$
\\
$\rm{B_c \to  K^+ \overline{K}^0}$ & $24.0^{+2.4}_{-0.0}(m_c)^{+7.3}_{-6.0}(a_i)^{+6.8}_{-5.8}(m_0)$&
$\rm{B_c \to K^+ \pi^0}$& $2.0^{+0.5}_{-0.3}(m_c)^{+1.2}_{-0.8}(a_i)^{+0.3}_{-0.1}(m_0)$
\\
\hline \hline
Decay Modes   &                & Decay Modes    &     \\
$(\Delta S=0)$& $BR's(10^{-7})$ & $(\Delta S=1)$ & $BR's(10^{-8})$ \\
\hline
 $\rm{B_c \to \pi^+ \rho^0}$
 &$1.7^{+0.1}_{-0.0}(m_c)^{+0.1}_{-0.2}(a_i)^{+0.6}_{-0.3}(m_0)$&
 $\rm{B_c \to K^+ \rho^0}$& $3.1^{+0.6}_{-0.8}(m_c)^{+1.2}_{-1.5}(a_i)^{+ 0.1}_{-0.2}(m_0)$  \\
 $\rm{B_c \to \overline{K}^0 K^{*+}}$
 &$1.8^{+0.7}_{-0.1}(m_c)^{+4.1}_{-2.1}(a_i)^{+0.1}_{-0.0}(m_0)$& $\rm{B_c \to K^0 \rho^+}$
 &$6.1^{+1.3}_{-1.5}(m_c)^{+2.5}_{-2.9}(a_i)^{+ 0.2}_{-0.3}(m_0)$\\
$\rm{B_c \to \pi^+ \omega}$ &
$5.8^{+1.4}_{-2.2}(m_c)^{+1.1}_{-1.3}(a_i)^{+ 0.4}_{-1.2}(m_0)$
& $\rm{B_c \to K^+ \omega}$ & $2.3^{+1.1}_{-0.3}(m_c)^{+1.8}_{-1.2}(a_i)\pm 0.1(m_0)$\\
\hline
 $\rm{B_c \to \rho^+ \pi^0}$  &$0.5^{+0.1}_{-0.1}(m_c)^{+0.3}_{-0.2}(a_i)^{+0.2}_{-0.3}(m_0)$&
$\rm{B_c \to K^{*0}\pi^+}$ &$3.3^{+0.7}_{-0.2}(m_c)^{+0.4}_{-0.4}(a_i)^{+0.2}_{-0.1}(m_0)$\\
 $\rm{B_c \to \rho^+ \eta}$ &$5.4^{+2.1}_{-1.2}(m_c)^{+0.9}_{-1.4}(a_i)\pm 0.0(m_0)$&
 $\rm{B_c \to K^{*+}\pi^0}$& $1.6^{+0.4}_{-0.1}(m_c)^{+0.3}_{-0.1}(a_i)^{+0.1}_{-0.0}(m_0)$ \\
$\rm{B_c \to \rho^+ \eta'}$ & $3.6^{+1.4}_{-0.8}(m_c)^{+0.6}_{-0.9}(a_i)\pm 0.0(m_0)$&
$\rm{B_c \to K^{*+} \eta}$ & $0.9^{+ 0.1}_{-0.0}(m_c) ^{+0.6}_{-0.2} (a_i)\pm 0.0(m_0)$\\
$\rm{B_c \to \overline{K}^{*0} K^+}$ & $10.0^{+0.5}_{-0.6}(m_c)^{+1.7}_{-3.3}(a_i)^{+0.0}_{-0.2}(m_0)$&
$\rm{B_c \to K^{*+} \eta'}$ & $3.8\pm 1.1 (m_c)^{+1.0}_{-0.6}(a_i)\pm 0.0 (m_0)$ \\
 & &  $\rm{B_c \to \phi K^+}$  & $5.6^{+1.1}_{-0.0}(m_c)^{+1.2}_{-0.9}(a_i)^{+0.3}_{-0.0}(m_0)$ \\
\hline \hline
\end{tabular}
\end{table*}



For $B_c \to VV$ decays, the decay rate can be written explicitly as,
\beq
\Gamma =\frac{G_{F}^{2}\bf{P_c}}{16 \pi m^{2}_{B_c} }
\sum_{\sigma=L,T}{\cal M}^{(\sigma)\dagger }{\cal M^{(\sigma)}}\;
\label{dr1}
\eeq
where $P_c\equiv |P_{2z}|=|P_{3z}|$ is the momentum of either of the outgoing vector mesons.
Based on the helicity amplitudes as defined in Eq.~(48) of Ref.~\cite{Liu2010:bcpv},
we can define the transverse amplitudes,
\beq
{\cal A}_{L}&=&-\xi m^{2}_{B_c}{\cal M}_{L}, \quad
{\cal A}_{\parallel}=\xi \sqrt{2}m^{2}_{B_c}{\cal M}_{N}, \non
&& {\cal A}_{\perp}=\xi m^{2}_{B_c} \sqrt{2(r^{2}-1)} {\cal M }_{T}\;. \label{eq:ase}
\eeq
for the longitudinal, parallel, and perpendicular polarizations,
respectively, with the normalization factor
$\xi=\sqrt{G^2_{F}{\bf{P_c}} /(16\pi m^2_{B_c}\Gamma)}$ and the
ratio $r=P_{2}\cdot P_{3}/(m_{M_1}\cdot m_{M_2})$.
These amplitudes satisfy the relation,
\beq
|{\cal A}_{L}|^2+|{\cal A}_{\parallel}|^2+|{\cal A}_{\perp}|^2=1
\eeq
following the summation in Eq.~(\ref{dr1}).

Since the transverse-helicity contributions manifest themselves in
polarization observables, we here define two kinds of
polarization observables, i.e., polarization fractions $(f_{L},f_{||},f_{\perp})$ and relative phases $(\phi_{||},\phi_{\perp})$ as
\beq
f_{L(||,\perp)}&=& \frac{|{\cal A}_{L(||,\perp)}|^2}{|{\cal A}_L|^2+|{\cal A}_{||}|^2+|{\cal A}_{\perp}|^2},\\
\phi_{||(\perp)} &\equiv& \arg \frac{A_{||(\perp)}}{A_{L}}\;;
\label{eq:pf}
\eeq
It should be noted that the final results of relative phases will plus one
value, i.e., $\pi$, due to an additional minus sign in the definition of ${\cal A}_L$.

In Table ~\ref{tab:bcvv}, we present the pQCD predictions for
CP-averaged branching ratios, the longitudinal polarization fractions ( $f_L's$)
and relative phases of the considered nine $B_c \to VV$ decays.
The dominant theoretical errors comes from the uncertainties of the charm quark mass
$\rm{m_c}=1.5\pm 0.15$ GeV, and the Gegenbauer moments $a_i$ of related meson distribution amplitudes, respectively.
The total error is the combination of individual errors in quadrature.

\begin{table}[thb]
\caption{The pQCD predictions of branching ratios(BRs), $f_L$, and the relative phases
$\phi_{||}$ and $\phi_{\perp}$ for $B_c \to V V$ decays.} \label{tab:bcvv}
\begin{tabular}[t]{l|llll} \hline  \hline
Decay Modes  & BRs($10^{-7}$)\hspace{0.3cm} & $f_L$(\%)\hspace{0.3cm} &$\phi_{||}$ (rad)\hspace{0.3cm} & $\phi_{\perp}$ (rad) \\
\hline
${\rm B_c \to \rho^+ \rho^0}$ &$0$ & $-$ &$-$ & $-$  \\
${\rm B_c \to \rho^+ \omega}$&$10.6^{+3.8}_{-0.3}$& $92.9^{+2.0}_{-0.1}$&$3.86^{+0.40}_{-0.32}$& $4.43^{+0.30}_{-0.25}$  \\
${\rm B_c \to \ov{K}^{*0} K^{*+}}$&$10.0^{+8.1}_{-4.8}$& $92.0^{+3.6}_{-7.1}$&$3.68^{+0.51}_{-0.25}$& $3.76^{+0.51}_{-0.20}$  \\
${\rm B_c \to K^{*0} \rho^+}$&$0.6^{+0.2}_{-0.1}$& $94.9^{+2.2}_{-1.5}$ &$4.11^{+0.34}_{-0.28}$  & $4.20^{+0.33}_{-0.22}$  \\
${\rm B_c \to K^{*+} \rho^0}$&$0.3^{+0.1}_{-0.1}$ & $94.9^{+1.4}_{-1.5}$ &$4.11^{+0.34}_{-0.28}$ & $4.20^{+0.33}_{-0.22}$  \\
${\rm B_c \to K^{*+} \omega}$&$0.3^{+0.0}_{-0.2}$ & $94.8^{+1.2}_{-1.3}$ &$4.15^{+0.28}_{-0.35}$ & $4.23^{+0.28}_{-0.26}$  \\
${\rm B_c \to \phi K^{*+}}$ &$0.5^{+0.1}_{-0.3}$  & $86.4^{+4.9}_{-9.1}$ &$3.80^{+0.51}_{-0.39}$ & $3.89^{+0.48}_{-0.28}$  \\ \hline \hline
\end{tabular}
\end{table}

From the numerical results in Table  \ref{tab:bcpp} , one can see that
\begin{enumerate}
\item[1]
Analogous to $B \to K \eta^{(\prime)}$ decays, the branching ratios of $B_c \to K \eta^{(')}$
modes also show a approximate relation:
\beq
Br(B_c \to K^+ \eta^\prime) \sim 10 \times Br(B_c \to K^+ \eta).
\eeq
This large difference can be understood by the destructive and constructive interference
between the $\eta_q$ and $\eta_s$ contribution to the $B_c \to K^+ \eta$ and
$B_c \to K^+ \eta^\prime$ decay.

\item[(2)]
The pQCD predictions for the branching ratios
vary in the range of $10^{-6} \sim 10^{-8}$~\cite{Liu2010:bcpv,Yang2010:bckk}, basically
agree with the predictions obtained by using the exact SU(3) flavor symmetry.
The $B_c\to \overline{K}^{*0} K^{+} $ and other decays with a decay rate
at $10^{-6}$ or larger could be measured at the LHC experiment.

\item[(3)]
For $B_c \to (\pi^+, \rho^+) (\eta, \eta')$ decays, the final state mesons $\eta$ and
$\eta^\prime$ contain the same component $\bar u u + \bar d d$, and the differences
among their branching ratios mainly come from the mixing
coefficients, i.e., $\cos\phi$ and $\sin\phi$.

\item[(4)]
Among the thirty $B_c$ decays considered in this subsection,  $B_c \to \rho^+\omega, \ov{K}^{*0} K^{*+}$
and $B_c\to \ov{K}^0K^{*+}$ have the largest branching ratios and are in the order of $10^{-6}$. This means that the annihilation contributions  to $B_c$ meson decays
may be rather important. We suggest the LHCb experiment to search for such decay modes.

\item[(5)]
For $B_c \to VV$ decays, the contributions coming from the longitudinal polarization play
the dominant role and the longitudinal polarization fractions $f_L\approx 95\%$,
except for $B_c \to \phi K^{*+}$ ( $ f_L \sim 86\%$). Unfortunately, it is very hard
to measure these decays, due to  the smallness of their decay rates:
in the range of $10^{-8}-10^{-7}$.

\end{enumerate}


\subsection*{ 5.3  Charmless hadronic $B_c\to SP,SV,AP,AV,AA$ decays} \label{ssec:anni-bc3}

For charmless $B_c \to SP$ and $SV$ decays, the pQCD predictions
for the branching ratios are in the range of $10^{-5}$ to $10^{-8}$~\cite{Liu2010:bcspv}.
Many decays with a decay rate at $10^{-6}$ or larger could be measured at
the LHCb experiment.
Similar to $B \to K^* \eta^{(\prime)}$ decays, the branching ratios
of $B_c \to \kappa \eta^{(')}$ channels also exhibit the interesting relation:
\beq
Br(B_c \to \kappa^+ \eta) \sim 5 \times Br(B_c \to \kappa^+ \eta').
\eeq
This difference can be understood by the destructive and constructive interference
between the $\eta_q$ and $\eta_s$ contribution to the
$B_c \to \kappa^+ \eta'$ and $B_c \to \kappa^+ \eta$ decay, respectively.

For $B_c \to K_0^*(1430) \eta$ and $B_c \to K_0^*(1430) \eta^\prime$ decays,
the pQCD predictions for their branching ratios ~\cite{Liu2010:bcspv}
are similar in size, since the factorizable contributions of $\eta_s$
term play the dominant
role. This feature will be tested by the near future experiments.
If $a_0$ and $\kappa$ are the $q\bar q$ bound states, the pQCD predictions
for $Br(B_c \to a_0 (\pi, \rho))$ and $Br(B_c \to \kappa K^{(*)})$
will be in the range of $10^{-6} \sim 10^{-5}$, which are within the reach
of the LHCb experiments and expected to be measured.

For the $a_0(1450)$ and $K_0^*(1430)$ channels, the branching ratios for
$B_c \to a_0(1450) (\pi, \rho)$ and $B_c \to K_0^*(1430) K^{(*)}$ modes
in the pQCD approach  are found to be of order $(5 \sim 47) \times 10^{-6}$
and $(0.7 \sim 36) \times 10^{-6}$ respectively ~\cite{Liu2010:bcspv}.
A measurement of them at the predicted level will
favor the structure of $q\bar q$ for the $a_0(1450)$ and
$K_0^*(1430)$ and identify which scenario is preferred.

For the pure annihilation $B_c \to AP, AV, AA$ decays, such as
\beq
B_c &\to& \bar{K}^0 (K_1(1270)^+, \quad K_1(1400)^+),  \quad a_1(1260)^+\omega,\non
&& b_1(1235) \rho,\quad (K_1(1270), K_1(1400)) K^*, \non
&& \rho^+ f_1(1285), \quad a_1(1260) b_1(1235), \non
&& (\bar{K}_1(1270)^0, \bar{K}_1(1400)^0) (K_1(1270)^+, K_1(1400)^+),\non
&& {\rm etc.}
\eeq
the pQCD predictions for their branching ratios are in the range
of $10^{-5}$ to $10^{-9}$~\cite{Liu2010:bcap,Liu2011:bcav,Xiao2011:bcaa}.
The decay modes with a sizable decay rate at $10^{-6}$ or
larger could be measured at the LHC experiments.

Since the QCD behavior of the $^1P_1$ meson is rather different from
that of the $^3P_1$ meson, the branching ratios in the pQCD approach of pure annihilation
$B_c \to A(^1P_1) (P, V, A(^1P_1))$ are basically larger than that of $B_c \to
A(^3P_1) (P, V, A(^3P_1))$ with a factor around $10 \sim 100$
\cite{Liu2010:bcap,Liu2011:bcav,Xiao2011:bcaa}, for example,
\beq
10 \times &&Br(B_c \to a_1(1260) (\pi, \rho)) \non
&&\approx Br(B_c \to b_1(1235) (\pi, \rho)),\non
100 \times && Br(B_c \to a_1(1260)^+ f_1(1285))\non
&&\approx Br(B_c \to b_1(1235)^+ h_1(1170)).
\eeq
These relations can be tested by the LHC experiments.

The pQCD predictions \cite{Liu2010:bcap,Xiao2011:bcaa} about the branching
ratios of some $B_c$ decays, such as
\beq
B_c &\to& K_1(1270)^+ \eta^{(\prime)}, K_1(1400)^+ \eta^{(\prime)},\non
&& K_1(1270)K, K_1(1400) K,  \bar{K}_1(1270)^0 K_1(1270)^+,\non
&& \bar{K}_1(1400)^0 K_1(1270)^+, \bar{K}_1(1270)^0 K_1(1400)^+,\non
&& \bar{K}_1(1400)^0 K_1(1400)^+,
\eeq
are rather sensitive to the value of the mixing angle $\theta_K$,
which will be tested by the running LHC experiments.
One can determine $\theta_K$ through the measurement of these decays if
enough $B_c$ events become available at the LHC experiments.
The pQCD predictions for several decays involving the mixtures
of $^3P_1$ and/or $^1P_1$ mesons are rather sensitive to the
values of the mixing angles, which can provide the important information
on both of sign and size of the mixing angles if they are detected
in the future experiments.
For $B_c \to VV, AV(VA), AA$ decays~\cite{Liu2011:bcav,Xiao2011:bcaa}
the longitudinal contributions play a dominant role in
most of those considered modes,  which will be tested by the ongoing LHC
experiments in the near future.

Once the above predictions on
the physical quantities in the pQCD approach can be
confirmed at the predicted level by the precision experimental
measurements in the future, which can also provide
indirect evidence for the important but controversial issues
(See Refs.~\cite{Arnesen08:anni-scet,Chay08:complexanni} for detail)
on the evaluation of annihilation
contributions at leading power, one can ask whether
it is almost real with a tiny strong phase in SCET or almost imaginary with a large strong
phase in the pQCD approach.

For all the considered charmless hadronic $B_c$ decays, the branching ratios of $\Delta S= 0$ processes are basically larger than those of $\Delta S =1$ ones.
Such differences are mainly induced  by
the CKM factors involved: $|V_{ud}|\sim 1 $ for the former decays
while $|V_{us}| \approx \lambda \sim 0.22$ for the latter ones.
Because only tree operators are involved, the {\it
CP}-violating asymmetries for these considered $B_c$ decays are absent naturally.

It should be stressed that the pQCD predictions still have large theoretical
uncertainties, mainly induced by the errors of the
Gegenbauer moments in the hadron distribution amplitudes and the errors of
decay constants $f_{B_c}$ and $f_{M_i}$.
By reducing these uncertainties, one can improve the
precision of the theoretical predictions effectively.
Moreover, only the short-distance contributions in the aforementioned hadronic $B_c$
decays are considered and perturbatively calculated by employing the pQCD approach.

The possible long-distance contributions to $B_c$ hadronic decays,
such as the rescattering effects, have been neglected in our calculations
since such contributions should be small based on the general expectations:
the perturbative contribution most possibly dominate the heavy $B_c$ meson decay.
One of the effective methods to decrease the theoretical error is to
define the ratios between branching ratios for suitable decay modes, such as those
connected through various $SU(3)$ flavor symmetries.

It is believed that all hadronic $B_c$ meson decays  will provide
important platform for studying the mechanism of annihilation
contributions, understanding the helicity
structure of the considered channels with vector and/or axial-vector meson(s)
and the content of the involved light scalar and axial-vector mesons.

\section*{6.  Summary and expectations } \label{sec:summary}

In the following years, the LHC experiments will collect more and more
$B_c$ production and decay events. The analysis of the huge number of
$B_c$ events do require precision theoretical predictions. On the other hand,
the properties of $B_c$ meson and many other different kinds of light or
heavy mesons, such as the relevant decay constants, the internal structure and
the mixing angles, etc., will be measured in the LHCb, CMS and ATLAS
experiments.

In this short review, we firstly summarize the recent progress of hadronic $B_c$ decays
at both experimental and theoretical aspects.
As aforementioned in Sec.~2, some hadronic $B_c$
decay channels have been detected by the LHCb experiment in the past four years.
The $B_c$ decays with a decay rate at the level of $10^{-6}$ can be detected
with a good precision at LHC experiments.

We then provide an outline about theoretical studies of
hadronic $B_c\to M_1 M_2$ decays with $M_i=(S,P,V,A,T)$
in the framework of the pQCD 
approach,
at the leading order and  leading power.
Up to now, about four hundred such decay modes have been studied in the pQCD approach,
some most important results are discussed here explicitly.
For more details of specific decays, one can see the original paper cited here.

For those considered $B_c$ decays, besides the emission diagrams,
the nonfactorizable spectator diagrams and the annihilation diagrams can also
be evaluated in the pQCD approach.
Furthermore, phenomenologically, it is found that the dominant contributions
to the branching ratios in many decay channels arise from
the nonfactorizable spectator and/or annihilation amplitudes. Such decay channels
can be classified into two sets:
(a) The $B_c$ decays with an emitted scalar or tensor meson, which cannot be
produced from the vacuum by the
(axial-)vector current or (pseudo-)scalar density in the SM;
and (b) the charmless hadronic $B_c$ meson decays, which can
only occur through the pure annihilation topology in the SM.

For the hadronic $B_c\to M_1 M_2$ decays considered in this short review,
the CP-averaged branching ratios vary in the range of $10^{-3} \sim 10^{-9}$.
Those decays with a decay rate at or larger than $10^{-6}$
can be measured in the near future LHC experiments.

For the considered annihilation dominant modes or pure annihilation
channels, the confirmation at the pQCD predictions through the precision
experimental measurements will provide  important information to the
controversial issues on how effectively and accurately to
evaluate the annihilation diagrams at leading power, which will provide more
important evidence on the sizable annihilation contributions
in heavy $B$ meson physics and further shed light on the
underlying mechanism of the annihilated $B$ meson decays.


\begin{acknowledgments}

This work is supported by the National Natural Science
Foundation of China under Grants No.~11205072, 10975074 and 11235005.

\end{acknowledgments}


\end{document}